\documentstyle[11pt,newpasp,twoside]{article}
\def\edcomment#1{\iffalse\marginpar{\raggedright\sl#1\/}\else\relax\fi}
\reversemarginpar

\begin{document}
\title{Once again about the origin of the system of the giant
stellar arcs in the Large Magellanic Cloud}

\author{Yu.N.Efremov}
\affil{Sternberg Astronomical Institute, Moscow State University,
Universitetskij pr. 13, 119992 Moscow, Russia}

\begin{abstract}
The origin of the arc-shaped stellar complexes in the LMC4
region is still unknown. These perfect arcs could not have been formed by
O-stars and SNe in their centers; the strong arguments exist also against
the possibility of their formation from infalling gas clouds. The origin
from microquasars/GRB jets is not excluded, because there is the strong
concentration of X-ray binaries in the same region and the massive old
cluster NGC 1978, probable site of formation of binaries with compact
components, is there also. The last possibility is that the source of energy
for formation of the stellar arcs and the LMC4 supershell might be the
the giant jet from the nucleus of the Milky Way, which might be active a
dozen Myr ago.
\end{abstract}

\section{Introduction}
In the field of the supershell LMC4 in north-east LMC a few huge arcs of young
stars and clusters have long time been known. The goal of this paper is to
turn attention to this unique system and still lasting enigma of its origin.

The brightest of these arcs was first noted by Westerlund and
Mathewson (1966), who wrongly identified as Shapley's
"Constellation III"; nowadays it is known as association LH77 or
"Quadrant", whereas the near-by smaller arc was called "Sextant"
(Efremov and Elmegreen 1998). These arcs are parts of exact
circles, they are the most perfect formations of a half of a
dozen more or less similar structures found inside other galaxies
(Efremov 2001a). So the LMC arcs are the best example of the
ordered stellar systems, whose regular appearance is surely not
due to gravitation. They are surely dynamically unstable.

The Quadrant arc is inside the HI superbubble LMC4, and Westerlund
and Mathewson (1966) suggested that both features were formed in
result of a Super-Supernova outburst. The whole system of the
arcs in this region was first noted by Hodge (1967); all in all
this area may host five arcs of young stars and clusters (see
Fig. 1 and 2).

The problem of origin of this system of arcs was first considered by Efremov
\& Elmegreen (1998). They found that age and radius of Quadrant are about 16
Myr and 280 pc, and of Sextant - 7 Myr and 170 pc, and suggested the formation
of the arcs from the gas shells, swept up by the sources of central pressure with
energy about $10^{52}$ ergs. The ages were derived from integral UBV photometry
by Bica et al. (1996). The Third arc is plausibly a by chance
configuration; it contains the clusters of different ages and the far
UV image, obtained by Smith et al. (1987) and presented in Fig.1b
supports this conclusion.

The HI superbubbles considered to have been formed by multiple super-
novae and O stars or by the impact of the high velocity clouds (Tenorio-Tagle
\& Bodenheimer, 1988). The supershells of swept up gas should break up to star
clusters located on an arc of a circle. However a lot of superbubbles is known
and only a few stellar arcs (Efremov 2001a; 2002). Only in IC 2574 galaxy
the HI supershell is known inside which there is an older cluster surrounded
by younger ones (Stewart \& Walter, 2000). However, even in this sample the
younger clusters are located in a random way, by no means in the regular arc
structure. Also, only 6 of 44 HI superbubbles in the Ho II galaxy contain clusters
suitable to be their progenitors (Rhode et al. 1999), yet even these supershells
are not surrounded by younger cluster arcs. It is possible that the regular arcs
of clusters were formed by other processes and because of this, they are so rare
objects.

\section{The LMC arcs were formed in a special way}

Apart from the general difficulties with explanation of the supershell origin, the
understanding of the LMC arcs faces to the specific problems. Efremov (1999)
and Efremov and Elmegreen (1999) have noted that it is not clear, why around
the clusters the very existence of which is doubtful, there could be so exotic
structures, missing around all known clusters elsewhere. Later on, Braun et al.
(2000) found that stellar population inside the LMC4, including the center of
Quadrant arc, have the age close to this arc age. These authors disproved also
the conclusion of Dopita et al. (1985) about the age gradient from the LMC4
center. It is not clear at all, why the giant arcs, so rare structures, are available
in the LMC in an amount not less than two (and probably five) and, moreover,
why all they are on a neighborhood with each other.

The Quadrant and Sextant arcs are parts of exact circles, instead
being the ellipses appropriate to inclination of the LMC plane to
the sky plane (Fig. 2b and 3). The latter should be the case
provided they are rings in the LMC disk plane. The most plausible
explanation of this is both arcs are cup-shaped, they are
segments of spherical surfaces, seen in projection (Fig. 4).

If the sources of the pressure were the SNe or O-stars and if the
gas density was uniform, the arcs must have shape of the rings,
not arcs or segments of spheres, because the young clusters, as
well as the highest gas density, are observed in the plane of the
symmetry of the galactic disks. For the case of the isotropic
pressure, the model experiments (Efremov et al. 1999) demonstrated
that the shape of the partial sphere may arise in two events: the
center of pressure is at outskirts of an isolated dense cloud, or
at some height outside the galactic plane. The latter position,
however, leads to the special orientation of the arc regarding
the line of intersection of the LMC and sky planes (Fig. 5). This
orientation contradicts to the observed ones of the LMC arcs,
apart from the Fifth arc only.

The special positions of progenitors of all arcs in dense isolated clouds being
improbable, they might be formed by the anisotropic pressure,
sources of which
may be located anywhere in or out the disk of the LMC. There exist however
other possibilities.

\section{High velocity clouds impact?}

The hypothesis about oblique infalling of group of high velocity
clouds could probably explain a neighborhood of all arcs, as well
as similar orientation of Quadrant and Sextant arcs. However it
is not the case for the Third and Fifth arcs. The different ages
of arcs are hardly compatible with this hypothesis, also. Only
the region near the SNR N49 (see Dopita et al. 1985, and
references therein) demonstrates the HI velocity perturbations,
which are seen also in Kim et al. (2003) data. The issue is how
long might exist the peculiarities in HI distribution and
velocities after a HVC impact. Anyway, this hypothesis is the
most natural explanation of absence of counterpart arcs.

The regular shape of the LMC arcs recalls the bow shock
appearance of the leading edge of some galaxies (say DDO 165 and
Ho II in M81 group) due to the ram pressure of the intergalactic
medium. Within framework of the HVC hypothesis one could suggest
that these clouds were dense enough and the stellar arcs were not
formed from the swept up gas shell, but are relics of the bow
shock at the interface of the LMC gas and the impacting HVC
(Efremov 2002). Earlier Braun et al. (2000) have proposed the ram
pressure from the Milky Way halo to explain the LMC4 supershell,
yet they have not suggested the mechanism of formation for the
stellar arcs.

The hypothesis of the HVC impact was applied by Comeron (2001) to the
explanation of the isolated stellar complex in M83, which is rather similar to
whole LMC4 region (Fig.6a), as was noted by Efremov (1999, 2001a). This M83
galaxy hosts also the second perfect arc of a few clusters (Fig. 6b) - again with
nothing in its center (Efremov 2001a).

\section{The GRB hypothesis}

As a source supplying $10^{52}$-$10^{53}$ ergs to the ISM and suitable to swept up
even the largest HI supershells, the Gamma-ray bursts (GRB) were suggested;
this idea was applied to the explanation of the LMC stellar arcs (Efremov 1999,
2001b and references therein). Their concentration in the same region of the
LMC was explained by ejection of the GRB progenitors in result of the dynamic
interaction of stars in the dense core of a star cluster. The unique massive 1 Gyr
old cluster is indeed there, it is NGC 1978, famous for its highly elliptical shape.

The pairs of compact objects, NS+NS or NS+BH might merge in GRB
event rather soon after ejection, at few hundred parsec distance
from the cluster. This hypothesis implies origin of binary stars
with compact components in result of dynamic interactions in
dense cores of rather old star clusters, what has been supported
now by many simulations of the dynamical evolution of clusters.
However, there are also data pointing to the connection of the
GRB with some type of the powerful Supernovae (Hypernovae). Their
energy is suffisiently large to form the stellar arcs, but
Hypernovae are suggested to have be originated from massive
stars, which should not exit in the old cluster, and escape of
which with rather high velocity is improbable.

The main argument for the Hypernovae origin of GRBs is that their
afterglow are observed mostly in regions of star formation. However, the GRB,
progenitors of which have being ejected from the same clusters might trigger
the star formation (like the LMC arcs) near-by. There are also independent evi-
dence for the origin of the short GRB from NS+NS or NS+BH merging (Perna,
Belczynski, 2002). Also, Bulik and Beclczynski (2003) found that NS+NS
progenitors of GRB must be short lived and traced star formation. On our opinion,
the distribution of GRB afterglows in a summary galaxy is close to that of the
classical old globular clusters and not to the regions of star formation. From the
distribution of GRB in redshift it follows that their average age was around a
few Gyrs, just compatible with the NGC 1978 age (Efremov 2001b).

However, the prevailing opinion now is that GRB energy is emitted within
the narrow superrelativistic jets whose energy is about only
$10^{51}$ erg. At any
rate, the only known in the LMC Soft repeating Gamma-ray source in the LMC
is also in the LMC4 area, within the SNR N49. The perturbations of HI velocities
just in this region might indicate the origin of this source in a Hypernovae event.

\section{The microquasars hypothesis}

There are other sources of relativistic jets which may be long
standing and whose accumulated energy is high enough.
Microquasars (MQS) are X-ray binaries which are able to emit
variable relativistic jets (Mirabel 2002); some 70\% of the X-ray
binaries in our Galaxy may be MQS (Fender and Maccarone 2003).
Their jets interact with the interstellar medium and can
therefore initiate star formation (Mirabel 2003). Also, for some
of them the high space velocities are observed, the examples
being GRO J1655-40, Cyg X-1, LS 5039 and Sco X-1; for the latter
the origin in a globular cluster was suggested (Mirabel and
Rodriques 2003).

The puzzling concentration of X-ray (XRB) binaries is known in
northern part of the LMC4 supershell, i.e. near NGC 1978. Half of
XRBs, known in the LMC, are there, within 3 square degrees of
total 80 square degree of the LMC area. The natural explanation
is again their formation in NGC 1978 and then ejection from this
cluster (Efremov 1999). Now many examples of connection between
XRB and the massive old clusters are known and many simulations
published, confirmed the possibility of XRB formation in and
ejection from dense old clusters.

There are little doubts that jets from MQS may trigger star
formation, providing they act long enough. The kinetic energy of
jets in SS433 - the most famous MQS in our Galaxy, is $10^{39}$
erg and during the life-time of about 20 000 years, the jets have
put in the interstellar medium $10^{51}$ ergs (Dubner et al.
1998); if the stage of the jet out similar to one observed now in
SS433 will be prolonged for some $10^{5}$ years, they could create
arcs similar to these in the LMC. The opening angle of the arcs
might be determined by the precession of jets, like known in
SS433, and also by its multiprecession, like suggested by
D.Fargion for GRB (see Efremov and Fargion 1999, and references
therein). However, the initially narrow jets may form arcs with
the large opening anwing to instabilities arising around a
relativistic jet. Moiseev et al. (2000) have put forward a
hypothesis that the cones of ionizations describing morphology of
regions, emanating narrow lines in a number of Seyfert galaxies,
are coupled to hydrodynamic instability caused by a velocity
break between the interstellar medium and a jet; the axes of
cones coincide with axes of jets, and the central angles at
centres of cones can make 50 - 60 degrees. We may accept that
pressure of the hot gas in such broad cone around of a narrow
jet, or the hot spot at its working surface are capable to
initiate star formation in a cup-shaped shell of swept up
interstellar gas. The similar situation may exist for jets of GRB
and/or of MQS.

\section{The jet from the Milky Way core?}

At any rate, the star formation in wider regions near tips of
relativistic jets is well known in galaxies Cen A and NGC 4258.
The sheet of contact of a jet working surface with interstellar
matter itself contains the hot spots which can be a power source
for creation of the expanding shell of the swept up gas, like
supergiant bubbles known around tips of very long jets in some
radio galaxies.

The very tip of the giant jet of radio galaxy 3C 445 has the
appearance of the arc about 9 kpc long - comparing with 450 kpc
long jet, and the synchrotron emission of the arc is seen also in
optics with the VLT as shown in Fig. 7 based on the paper by
Prieto et al. (2002). If the similar ratio is fair for the LMC
case, it could mean, that the sources of jets generated the
arc-shaped stellar complexes might be at dozens kiloparsecs from
the LMC. Let us note again that the orientation of two most
expressed and exact arcs is similar to each other and to the
older Fourth arc.

The hypothesis may be advanced that the the source of the jet was the
nucleus of our Galaxy. It is not active now, yet various signs of its past activity
are long time suspected. Long jets ejecting under rather small angle to the
galaxy plane are known in some spiral galaxies; Gopal-Krishna and Irwin (2000)
believed that such jets may give origin to the expanding shells, especially in the
region of the fast decrease of the galaxy gas density. It is known also that the
precessing long jet of NGC 4258 galaxy was sometime ago in the plane of the
galaxy (Cecil et al. 2000).

All in all, existing data does not contradict to suggestion that some 15 Myr
ago the Galaxy emitted the long relativistic jet which might be directed to the
LMC (Fig. 8). It might form the LMC4 HI superbubble whereas the hot spots
within it might give rise to smaller shells, relics of which are observed now as
the giant stellar arcs. The ratio between the jet length and the size of the arc at
its tip is 50 for the 3C 445 case, and it is about the same between the distance
to the LMC and size of the LMC4 supershell. Note the exceptional nature of
the LMC4 superbubble which is the largest in the LMC and is about completely
empty from the gas, unlike any other supershells in the LMC. The position of
LMC4 at outskirts of the LMC HI disk consists with mechanism of supershell
formation by a galactic jet  at the gas density gradient, suggested by
Gopal-Krishna and Irwin (2000).

Anyway, the angle between the MW center and the plane of the LMC, seen
from the center of the LMC, is now about 63 degree (van der Marel 2001).
Note also that the direction of the LMC proper motion to ENE is compatible
with the relative ages and positions of the stellar arcs, the oldest Fourth
arc being at the East and youngest Sextant being at the West.
The slow precession
motion of the suggested jet might be also a reason for this correlation of ages
and positions. Then the LMC4 superbubble might be a result of the pressure
from the Quadrant stars, as Efremov and Elmegreen (1998) have suggested.
The hypothesis on the origin of the arcs under action of the past jet from
the Milky Way nucleus does not explain the presence in the same region of the
LMC the excess density of X-ray binaries and the unusual cluster, but does
connect the stellar arcs and the LMC4 superbubble origin. Also, the jet from a
neighbor large galaxy core might explain the exceptional nature of the LMC arc
system. This hypothesis may be crazy enough to be correct. More data on the
past history of the Milky Way nuclear activity and surely on the LMC4 region
are needed to check it.
\vspace{0.5cm}

{\bf Acknowledgement}

\vspace{0.3cm}

Thanks are due to A.Chernin, B.Elmegreen,
P.Hodge and J.Palous for many discussions concerning the LMC arcs.
I am grateful to E.Yu.Efremov for modeling the Quadrant and
Sextant arcs, as shown in Fig. 4.

The support from the RFBR (project 03-02-16288) and Scientific
School Foundation (project NSh 389.2003.2) is appreciated.

\vspace{0.5cm}

{\bf Figure captions}

\vspace{0.3cm}

{\bf Fig. 1.}

a) The LMC in the blue band. The LMC4 region is in the
upper left corner.

b) The LMC in the vacuum UV (the effective
band-pass 1500 A). Only the hot young stars are seen. The image
is the edited Plate 22 from Smith et al. (1987).

{\bf Fig. 2.} The supershell LMC4 region in the LMC.

a) The edited
image of the plate, obtained by H.Shapley at Boyden observatory,
courtesy by P.Hodge.

b) The stellar arcs stressed by rings. The
brightest parts are white. 1 - Quadrant, 2 - Sextant, 3 - Third
arc (plausibly unreal, compare to Fig. 1b), 4 - Fourth arc, 5 -
Fifth arc, 6 - NGC 1978 cluster, 7 - SN remnant N49.

c) The HI distribution, based on Kim et al. (2003) data, and the
star clusters from Bica et al. (1996) catalog of UBV integral
photometry. The clusters within Quadrant and Sextant arcs are
stressed. The NGC 1978 cluster and the SN remnant N49 are shown as
squares.

{\bf Fig. 3.} Sextant arc as seen by UIT (the image from APOD),
with the regular ring overlaid.

{\bf Fig. 4.}

a) Quadrant arc (top) and its model (bottom) in the form a spherical
circle of angular diameter 130 degrees if viewed from the center
of the sphere, with the symmetry axis at 80 degrees to the sky
plane and with a thickness-to-radius ratio of 0.18.
A total of 1200 points are inside the segment of sphere and
4000 field points are used.

b) Sextant arc (top, UIT image) and its model
(bottom) in the form a spherical circle of angular diameter 80
degrees if viewed from the center of the sphere, with the symmetry
axis at 80 degrees to the sky plane and with a thickness-to-radius
ratio of 0.10. A total of 300 points are inside the segment of sphere and
1000 field points are used.

{\bf Fig. 5.} The projected column density of shells from
isotropic pressure source at 100 pc above (left panel) and 100
below (right panel) the LMC plane. The LMC line of nodes is
shown. The 300 pc long arrow points toward the center of the LMC.
It is the Fig. 6 of Efremov, Ehlerova and Palous (1999).

{\bf Fig. 6.}

a)The isolated stellar complex in M83, which possibly includes
two stellar arcs, the edited VLT image (Efremov 2001a).

b)The arc of stellar clusters at the West of M83, the edited VLT
image (Efremov, 2001a).

{\bf Fig. 7.} The synchrotron emission of jets from radio galaxy
3C 445 (Prieto et al., 2002; the fig. is from ESO Press release).

{\bf Fig. 8.} The Milky Way and the Magellanic Clouds in the color
composite photography. The LMC4 region is seen to the right of
the upper tip of the LMC bar. The edited image from APOD.
\end{document}